\documentclass{amsart}
 \usepackage{amsaddr}
\usepackage[english]{babel}

\usepackage[letterpaper,top=2cm,bottom=2cm,left=3cm,right=3cm,marginparwidth=1.75cm]{geometry}

\usepackage{amsmath}
\usepackage{graphicx}
\usepackage[colorlinks=true, allcolors=blue]{hyperref}

\newtheorem{thm}{Theorem}
\newtheorem{prop}[thm]{Proposition}
\newtheorem{lem}[thm]{Lemma}
\newtheorem{cor}[thm]{Corollary}
\newtheorem{defn}[thm]{Definition}

\newtheorem{rmk}[thm]{Remark}
\newenvironment{pf}{\begin{proof}}{\end{proof}}

\newcommand\ZZ{\mathbb Z}

\newcommand{\PN}{\textrm{PN}}
\newcommand{\ILPN}{\textrm{ILPN}}

\title{A polynomial invariant for a new class of phylogenetic networks} 

\author{Joan Carles Pons}
\address{Department of Mathematics and Computer Science,  University of the Balearic Islands, E-07122 Palma, Spain}
\email{joancarles.pons@uib.es}
\author{Tomás M. Coronado}
\address{Department of Mathematics and Computer Science,  University of the Balearic Islands, E-07122 Palma, Spain}

\author{Michael Hendriksen}
\address{School of Mathematics and Statistics, University of Melbourne, Australia}
\author{Andrew Francis}
\address{Centre for Research in Mathematics and Data Science,  Western Sydney University, Parramatta, Australia}

\begin{document}
\maketitle

\begin{abstract}
Invariants for complicated objects such as those arising in phylogenetics, whether they are invariants as matrices, polynomials, or other mathematical structures, are important tools for distinguishing and working with such objects. 
In this paper, we generalize a complete polynomial  invariant on trees to a class of phylogenetic networks called separable networks, which will include orchard networks. Networks are becoming increasingly important for their ability to represent reticulation events, such as hybridization, in evolutionary history.  
We provide a function from the space of internally multi-labelled phylogenetic networks, a more generic graph structure than phylogenetic networks where the reticulations are also labelled,
 to a polynomial ring. We prove that the separability condition allows us to characterize, \textit{via} the polynomial, the phylogenetic networks with the same number of leaves and same number of reticulations by considering their internally labelled versions.
 While the invariant for trees is a polynomial in $\mathbb{Z}[x_1,\ldots, x_n,y]$ where $n$ is the number of leaves, the invariant for internally multi-labelled phylogenetic networks is an element of $\mathbb{Z}[x_1,\ldots, x_n,\lambda_1,\dots,\lambda_r,y]$, where $r$ is the number of reticulations in the network.  When the networks are considered without leaf labels the number of variables reduces to $r+2$.
\end{abstract}

\section{Introduction}

A complete polynomial  invariant able to uniquely distinguish between rooted trees has been recently introduced in \cite{liu2021tree}. Motivated to analyze and compare tree shapes in a phylogenetic context, this polynomial (to which we will refer as the \emph{Liu polynomial}) has been used both to define a similarity measure on rooted tree shapes and to estimate parameters and models \textit{via} its coefficients \cite{liu2020polynomial}. Moreover, its generalization from trees to networks (by analyzing the set of embedded spanning trees in the network) has also been used to study the properties of randomly generated networks  \cite{janssen2021comparing}. 

We note that the word ``invariant'' is used here in its traditional sense, and not the one used in algebraic geometry approaches to phylogenetics, in which phylogenetic invariants for an evolutionary model along a tree are the polynomials which vanish on the expected frequencies of base patterns at the leaves \cite{cavender1987invariants}.
Throughout this article,  a \emph{(complete) invariant} of a set $A$ is a function $f:A\to B$ with the property that $x\sim_A y$  if and only if $f(x)\sim_B f(y)$, where $B$ is some other set (such as the set of polynomials), and $\sim_A$ and $\sim_B$ are equivalence relations in the respective sets. 

A multitude of (non-polynomial) invariants have been defined for specific subclasses of phylogenetic networks. To name just a few, the $\mu$-vectors which store the number of paths from nodes to leaves characterize (among others) tree-child networks \cite{cardona2008comparison} and orchard networks (without stacks) \cite{bai2021defining};  the set of displayed trees characterizes regular networks \cite{willson2010regular}; and the induced trinets (minimal subnetworks induced by triples of leaves) that characterize (among others)  level-$2$ networks \cite{van2014trinets} and orchard networks \cite{semple2021trinets}.

In this paper we show how a polynomial invariant can be defined for rooted phylogenetic networks,
generalizing the Liu polynomial invariant for trees. 
In order to do so, we consider phylogenetic networks and a labelled version of them, called internally labelled phylogenetic networks, where we keep the labels on leaves and also (bijectively) label  the reticulations. In fact, internally labelled phylogenetic networks are a subset of a more general set of networks, which we call internally multi-labelled phylogenetic networks, or IMLN's. On these networks the presence of elementary nodes is allowed, and leaves, reticulation and elementary nodes are all labelled.  Then, if we denote by $\PN$ the set of all phylogenetic networks (up to isomorphism) and by $\ILPN$ the set of all internally labelled phylogenetic networks (up to isomorphism), the map $\Phi:\ILPN\to\PN$ that sends each internally labelled phylogenetic network to the phylogenetic network obtained by ``forgetting'' all the internal labels (on reticulations) is obviously well defined; therefore for each $N\in\PN$, $\Phi^{-1}(N)$ is the set of all the internally labelled phylogenetic networks that have its same topology; its fiber, in mathematical terms.
The aim of this paper is to define a polynomial $p$ that uniquely characterizes these fibers and, in so doing, also characterizes the phylogenetic networks beneath them. 



This paper is organized as follows. In the Definitions section we include the three main graph structures of study: phylogenetic networks, internally labelled phylogenetic networks and internally multi-labelled phylogenetic networks (or IMLN's). We also define the concept of isomorphism on these structures. The Folding and unfolding section studies a process that unfolds an IMLN into a tree (an IMLT) and its reverse, folding, that recovers the initial IMLN. The key result of this section is the characterization of an IMLN by an IMLT  (Corollary~\ref{cor:fu}). The next section is dedicated to the definition and study of an extension of the Liu polynomial on IMLN's. If $N$ is an IMLN on a set of leaves labelled by $X$, the assigned polynomial $p(N)$  has  $|X|+r+1$ variables, where $r$ is the number of reticulations in the network.  This section is further divided into multiple subsections. The first one studies a special type of path (composed only of reticulations or elementary nodes) in IMLN's, called strong paths.  Roughly speaking, these allow us to define an equivalence relation between IMLN's, and we prove that two IMLN's share the polynomial if, and only if, they are equivalent (Theorem~\ref{thm:imlt-inj}). The second gives a sufficient condition on the space of phylogenetic networks (which we call separability) for the derived internally labelled phylogenetic networks to be completely characterized by the polynomial. The multiple lemmas proved in this part allow us to prove the main result (Theorem~\ref{teo:pol-iso}) in the third part; that is, the polynomial is a complete invariant in the set of internally labelled separable phylogenetic networks  up to isomorphism. The fourth subsection proves that orchard networks are separable, and so are characterized by the polynomial introduced in this paper (Theorem~\ref{thm:sep-orch}). Finally, in the last part,  we present how the obtained results can be applied for an unlabelled version of networks, in the sense that we forget the labelling of the leaves, reducing the polynomial to $r+2$ variables (Proposition~\ref{prop:unlabelled}). 

\section{Definitions}
In this section we introduce the mathematical notation that will be used in the rest of the paper.

Throughout this paper, $X$ will denote a non-empty finite set (of taxa). Commonly, we will use $X=\{x_1,\ldots, x_n\}$, and we will allow ourselves to see each member of $X$ as an irreducible polynomial in $\ZZ[x_1,\ldots,x_n]$; i.e., we will consider the labels of the leaves in our networks to be polynomials of the form $x_i$ for $i\in\{1,\ldots, n\}$.

\begin{defn}\label{d:rooted.binary.net}
A \emph{rooted binary phylogenetic network $N=(V,E)$ on $X$}, or simply a \emph{phylogenetic network} on $X$, is a rooted directed acyclic graph with no parallel arcs satisfying the following conditions:
\begin{enumerate}
    \item[(i)] any node with out-degree zero (a \emph{leaf}) has in-degree  one, and the set of nodes with out-degree zero, denoted by $L(N)$, is identified with $X$ \textit{via} a bijection $\varphi: L(N)\to X$;
    \item[(ii)] the root is the only node with in-degree zero, and has out-degree two;
    \item[(iii)] any other node has either in-degree one and out-degree two (a \emph{tree} node), or in-degree two and out-degree one (called a \emph{reticulation} node).
\end{enumerate}
We shall consider the leaves and root to be tree nodes.
\end{defn}

\begin{defn}\label{d:IMLN}
A \emph{rooted binary internally multi-labelled phylogenetic network $N=(V,E)$ on $X$}, or simply an \emph{IMLN} on $X$, is a rooted directed acyclic graph with no parallel arcs satisfying the following conditions:
\begin{enumerate}
    \item[(i)] any node with out-degree zero (a \emph{leaf}) has in-degree one, and the set of nodes with out-degree zero, denoted by $L(N)$, is identified with $X$ \textit{via} a surjection $\varphi: L(N)\to X$; 
    \item[(ii)] the root is the only node with in-degree zero, and it can have out-degree one (in which case we shall say it is an \emph{elementary} node) or two (a \emph{tree} node);
    \item[(iii)] any other node has either in-degree one and out-degree two (again, a \emph{tree} node), or in-degree two and out-degree one (called a \emph{reticulation} node), or in-degree one and out-degree one (again, an \emph{elementary} node);
    \item[(iv)] if $R(N)$ denotes the set of reticulation nodes and $E(N)$ the set of elementary nodes of $N$, then there exists $\ell:R(N) \cup E(N)\to \{\lambda_1,\ldots, \lambda_r\}$ a labelling function such that its restriction to $R(N)$ is injective  and if $u\in R(N)$ and $v\in E(N)$, $\ell(u)\neq \ell(v)$.
\end{enumerate}
\end{defn}

\begin{defn}
A \emph{rooted binary internally multi-labelled phylogenetic tree $T=(V,E)$ on $X$}, or simply \emph{IMLT} on $X$, is an IMLN without reticulation nodes. 
\end{defn}

We will consider the labels $\lambda_1,\ldots,\lambda_r$ to be irreducible polynomials in $\ZZ[x_1,\ldots,x_n,\lambda_1,\ldots,\lambda_r]$. Notice that Definition~\ref{d:IMLN} implies that IMLN's are a recursive structure in the following sense: given any IMLN $N$, for any $u\in V(N)$, the subgraph rooted at $u$ is still an IMLN. This is not the case in general for phylogenetic networks. 

In the case that an IMLN (with the root of out-degree two) does not have elementary nodes and the labelling on the leaves is a bijection, by definition, it becomes a phylogenetic network if the labelling $\ell$ on reticulations is suppressed. Also, if we consider a phylogenetic network and we add a labelling bijection $\ell:R(N)\to \{\lambda_1,\ldots, \lambda_r\}$, it becomes an IMLN. In order to reflect this possibility, we introduce the following definition.

\begin{defn}
An \emph{internally labelled phylogenetic network} $N$ on $X$ is an IMLN on $X$ without elementary nodes and where the maps $\varphi: L(N)\to X$ and $\ell:R(N)\to \{\lambda_1,\ldots, \lambda_r\}$ are bijections. 
\end{defn}

In order to formally define the concept of isomorphism between a pair of phylogenetic networks or between a pair of IMLN's, we consider the alternative notation, $(V,E,\varphi)$ and $(V,E,\varphi, \ell)$, to reflect the labelling functions, respectively.
\begin{defn}
Two phylogenetic networks $N_1=(V_1,E_1,\varphi_1)$ and $N_2=(V_2,E_2,\varphi_2)$ on $X$ are \emph{isomorphic} if there exists a bijection $f: V_1 \rightarrow V_2$ such that 
$\varphi_1(x)=\varphi_2(f(x))$ for all $x \in L(N_1)$,
and $(u,v) \in E_1$ if and only if $(f(u), f(v)) \in E_2$.
\end{defn}
\begin{defn}
Two IMLN's $N_1=(V_1,E_1,\varphi_1,\ell_1)$ and $N_2=(V_2,E_2,\varphi_2,\ell_2)$ on $X$ are \emph{isomorphic} if there exists a bijection $f: V_1 \rightarrow V_2$ such that
$\varphi_1(x)=\varphi_2(f(x))$ for all $x \in L(N_1)$, $\ell_1(x)=\ell_2(f(x))$ for all $x \in R(N_1) \cup E(N_1)$,
and $(u,v) \in E_1$ if and only if $(f(u), f(v)) \in E_2$.
\end{defn}

That is, a graph isomorphism that preserves the labels of both the reticulation and elementary nodes.

\section{Folding and unfolding}\label{sec:fold}

Following \cite{huber2016folding}, a phylogenetic network can be ``unfolded" in a specific manner to obtain a multi-labelled tree, that is a particular IMLT without elementary nodes in terms of the previous definitions. Moreover, in some cases, this process can be reverted, and the multi-labelled tree can be ``folded'' recovering the initial network. A phylogenetic network cannot in general be characterized by a multi-labelled tree, and this correspondence is valid only for the subclass of $FU$-stable phylogenetic networks \cite{huber2016folding}.

In this section, however, we prove that 
an internally labelled phylogenetic network
can be uniquely characterized by an IMLT obtained by a sequence of ``unfoldings'' on its reticulation nodes. Roughly speaking, considering the reticulations of an IMLN in a specific order, it is possible to sequentially duplicate the subnetwork descending from these nodes until an IMLT is obtained.

Let $N$ be a (generic) IMLN, and $R(N)$ the set of its reticulation nodes. The relation of being a descendant of another node induces a partial order over $R(N)$, which we will denote by $\leq_R$. That is, for any two nodes $u,v \in R(N)$, $u\leq_R v$ if, and only if, there exists a directed path from $v$ to $u$. Let $R_{\min}(N)$ be the set of the minimal elements of $R(N)$ under this order, i.e. reticulation nodes such that none of their descendants are also reticulation nodes. 

\begin{lem}\label{lema:min_tree}
Let $N$ be an IMLN and $u \in R_{\min}(N)$. Then the graph rooted at $u$ is an IMLT. 
\end{lem}
\begin{pf}
If $u \in R_{\min}(N)$, then there is no path in $N$ from $u$ to another reticulation. This means that there are no reticulations in the graph rooted at $u$; and therefore it is an IMLT. 
\end{pf}

Let $N$ be an IMLN, and consider $u\in R_{\min}(N)$ (so that $u$ is labelled by an element in $\{\lambda_1, \ldots, \lambda_r\}$). Let $v_1,v_2$ be its parents, noting that $v_1\neq v_2$ due to the fact that parallel arcs are excluded. Define $U(N, u)$ to be the \textit{unfolded IMLN of $N$ at $u$}, obtained by the following algorithm:
\begin{enumerate}
		\item delete edges $(v_1,u)$ and $(v_2,u)$;
		\item duplicate $N(u)$, the IMLT rooted at $u$, including all its labels; 
		\item add an edge from $v_1$ to one of the resulting copies of $u$, and an edge from $v_2$ to the remaining copy of $u$.
\end{enumerate}

\begin{rmk}\label{rem:unfolded-paths}
Notice that the process of unfolding preserves paths in the following sense: if $N'$ is obtained from $N$ by unfolding $N$ at some node $u$, then any path between two nodes in $N'$ comes from an existing path in $N$; and \textit{vice versa}, any path between two nodes in $N$ corresponds to a path in $N'$. Notice, however, that a path in $N$ might very well correspond to two different paths in $N'$, and so this assignation is not injective.
\end{rmk}

\begin{cor}
Let $N$ be an IMLN, and $u\in R_{\min}(N)$. Then $U(N, u)$ is an IMLN. 
\end{cor}

Let $N$ be an IMLN. We say that a sequence $(u_1,\ldots, u_k)$ of nodes in $R(N)$
is \emph{compatible} if the associated sequence $(N, N_{u_1}, N_{u_2},\ldots, N_{u_k})$ of IMLN's, such that $N_{u_{i+1}} = U(N_{u_i}, u_{i+1})$ and $N_{u_1}=U(N,u_1)$, is such that $u_{i+1}\in R_{\min}(N_{u_i})$ and $u_{1}\in R_{\min}(N)$. Then, if $(u_1,\ldots, u_k)$ 
is compatible,  for each $i \in \{1,\ldots,k-1\}$ there is no path from $u_i$ to $u_j$ when $j>i$; i.e., it is non decreasing under the partial order $\leq_R$ induced by the network over $R(N)$.

\begin{lem}\label{lem:unf-order}
Let $N$ be an IMLN and $u_1, u_2\in R_{\min}(N)$. Then,
$$
U(U(N, u_1), u_2) = U(U(N, u_2), u_1).
$$
\end{lem}
\begin{pf}
It is straightforward by Lemma \ref{lema:min_tree} and the steps of the unfolding algorithm. If $u_1 \in R_{\min}(N)$, then $ u_2\in R_{\min}(U(N,u_1))$; otherwise there would be a reticulation node $u'$ in $R(U(N,u_1))$ and a path from $u_2$ to $u'$ in $U(N,u_1)$, and so in $N$, which is a contradiction. Then, by Lemma \ref{lema:min_tree}, the graph rooted at $u_2$ in $U(N,u_1)$ is an IMLT. Since $u_2$ is not a node in any of the copies of the IMLT rooted at $u_1$ in the construction of $U(N,u_1)$, there is no intersection between the copies from $u_1$ and the copies from $u_2$. Since the same argument holds if we start by $u_2$, the result is achieved.
\end{pf}

Lemma \ref{lem:unf-order} can be extended following the same arguments for any set of reticulations $\{u_1,\ldots, u_k\}$ if all of them are in $R_{\min}(N)$, since there will be no intersection between the created copies of IMLT's.

Let $N$ be an IMLN. We define an equivalence relation $\equiv$ in the set of compatible sequences of elements of $R(N)$ as follows: $$ (u_1, u_2, \ldots, u_k) \equiv  (v_1, v_2, \ldots, v_{k'}) \Leftrightarrow \{u_1, u_2, \ldots, u_k\}= \{v_1, v_2, \ldots, v_{k'}\}.$$ That is, we say that two compatible sequences are \emph{equivalent} if they are composed by the same set of nodes.

An \emph{$\leq_R$-chain} in an IMLN $N$ is a chain  under the $\leq_R$ order defined on $R(N)$ (or a subset of it). That is, a subset of reticulations such that $u_1 \leq_R \cdots \leq_R u_s$. And, an \emph{$\leq_R$-antichain} in an IMLN $N$ is an antichain under the $\leq_R$ order; i.e., a subset of reticulations of $N$ which are pairwise incompatible ($u_i \not\leq_R u_j$ and $u_j \not\leq_R u_i$ if $u_i \neq u_j$) under the $\leq_R$ order.

In the next lemma we prove that if we consider an $\leq_R$-chain in an IMLN $N$ then there is a single way to traverse these nodes in a  compatible sequence, from bottom to top. On the other hand, if we consider an $\leq_R$-antichain, then every way to traverse these nodes is valid to form a compatible sequence.

\begin{lem}\label{lem:chains}
Let $N$ be an IMLN and $S=\{v_1,\ldots, v_r\} \subseteq R(N)$. Then
\begin{itemize}
    \item[(a)] If $v_1 \leq_R v_2 \leq_R \cdots \leq_R v_r$ is an $\leq_R$-chain, then $v_i$ must precede $v_j$ in every compatible sequence containing $S$ if $i<j$.  
    \item[(b)] If $S$ is an $\leq_R$-antichain, then every possible ordering of its nodes produces a compatible sequence composed by $S$.
\end{itemize}
\end{lem}
\begin{pf}
We first prove (a). If $v_1 \leq_R v_2 \leq_R \cdots \leq_R v_r$ is an $\leq_R$-chain, then there is a path from $v_j$ to $v_i$ if  $i<j$. Therefore if there exists a path from $v_i$ to $v_j$, it produces a cycle in $N$; but this is not possible because $N$ is an IMLN, and so in particular it is acyclic. This means that there is no path from $v_i$ to $v_j$ when $i<j$. Consequently,  if $i<j$, $v_i$ must precede $v_j$ in every compatible sequence containing $S$. 
\\
Now we prove (b). Let $v$ and $v'$ be two nodes in $S$. If $v$ precedes $v'$ in a sequence there cannot be a path from $v$ to $v'$; otherwise $v'\leq_Rv$. If $v'$ precedes $v$ in a sequence there cannot be a path from $v'$ to $v$; otherwise $v\leq_Rv'$. Since $S$ is an $\leq_R$-antichain, then both cases derive compatible sequences. 
\end{pf}

\begin{cor}\label{cor:unf-order}
Let $N$ be an IMLN and $(u_1, u_2, \ldots, u_k) \equiv (v_1, v_2, \ldots, v_{k})$ a pair of equivalent compatible sequences of elements of $R(N)$. Let $(N, N_{u_1}, N_{u_2},\ldots, N_{u_k})$ and $(N, N'_{v_1}, N'_{v_2},\ldots, N'_{v_k})$ be the associated sequences of IMLN's to their corresponding compatible sequences. Then  $N_{u_k}$ and $N'_{v_k}$ are isomorphic.
\end{cor}
\begin{pf}
For $k=1$ there is nothing to prove, since $u_1=v_1$.
 For $k=2$. If $u_1,u_2 \in R_{\min}(N)$, there is nothing to prove, because $(u_1,u_2)$ and $(u_2,u_1)$ are compatible sequences and Lemma \ref{lem:unf-order} applies. If $(u_1,u_2)$ is a compatible sequence and $u_1 \leq_R u_2$, then must be $(v_1,v_2)=(u_1, u_2)$ (and not $(v_1,v_2)=(u_2, u_1)$), since $u_1\notin R_{\min}(N_{u_2})$.
\\
The general situation for $k\geq 3$ demands a different approach. Let $s_1=(u_1, u_2, \ldots, u_k)$ and $s_2=(v_1, v_2, \ldots, v_{k})$. Since $s_1 \equiv s_2$, we have $\{u_1, u_2, \ldots, u_k\}= \{v_1, v_2, \ldots, v_{k'}\} \subseteq R(N)$. Let $A=\{u_1, u_2, \ldots, u_k\}$. Then we could iteratively apply the following process to prove the result. Let $A'=\{u\in A: u \in R_{\min}(N)\}$. Note that $A'$ is not empty due to $u_1$ and $v_1$ (which could be equal) are in $R_{\min}(N)$. Then, let $s_1^{A'}$ be the sequence obtained from $s_1$ by moving all the nodes in $A'$ to the first positions (in such a way that if $u_i, u_j \in A'$ with $i<j$, then the node $u_i$ appears before $u_j$ in $s_1^{A'}$) and remain invariant the rest of nodes. Note that $s_1^{A'}$ is compatible by construction and $s_1^{A'} \equiv s_1$.  A similar  process can be repeated to obtain $s_2^{A'} \equiv s_2$. Note that the set of nodes  of $A'$ occupying the first $|A'|$ positions in both $s_1^{A'}$ and $s_2^{A'}$ are exactly the same, and it is an $\leq_R$-antichain;  but these nodes may not appear in the same order in both sequences.
\\
Let $u^*$ be the last node (the rightmost) in $s_1^{A'}$ such that $u^* \in A'$. Now let $\hat{s_2}^{A'}$ be the compatible equivalent sequence to $s_2^{A'}$ obtained by remaining invariant all positions except for the node $u^*$, which comes to be the last node in $\hat{s_2}^{A'}$ with $u^* \in A'$. This ensures that the last node of the first $|A'|$ positions in both $s_1^{A'}$ and $\hat{s_2}^{A'}$ is the same, $u^*$. Note that, could be $u^*=u_k=v_k$ (when $A=A'$).  By Lemma \ref{lem:chains}(b) and Lemma \ref{lem:unf-order}, the IMLN $N_{u^*}$ obtained by sequentially unfold at the nodes in $s_1^{A'}$ until $u^*$ is achieved, is isomorphic to the IMLN obtained by sequentially unfold at the (same) nodes in $\hat{s_2}^{A'}$ until $u^*$ is achieved. Then, the same process can be repeated by considering new equivalent compatible sequences obtained from  $s_1^{A'}$ and $\hat{s_2}^{A'}$ by suppressing the first $|A'|$ positions and starting with the IMLN $N_{u^*}$.\end{pf}

\begin{figure}[!ht]
\includegraphics[width=\linewidth]{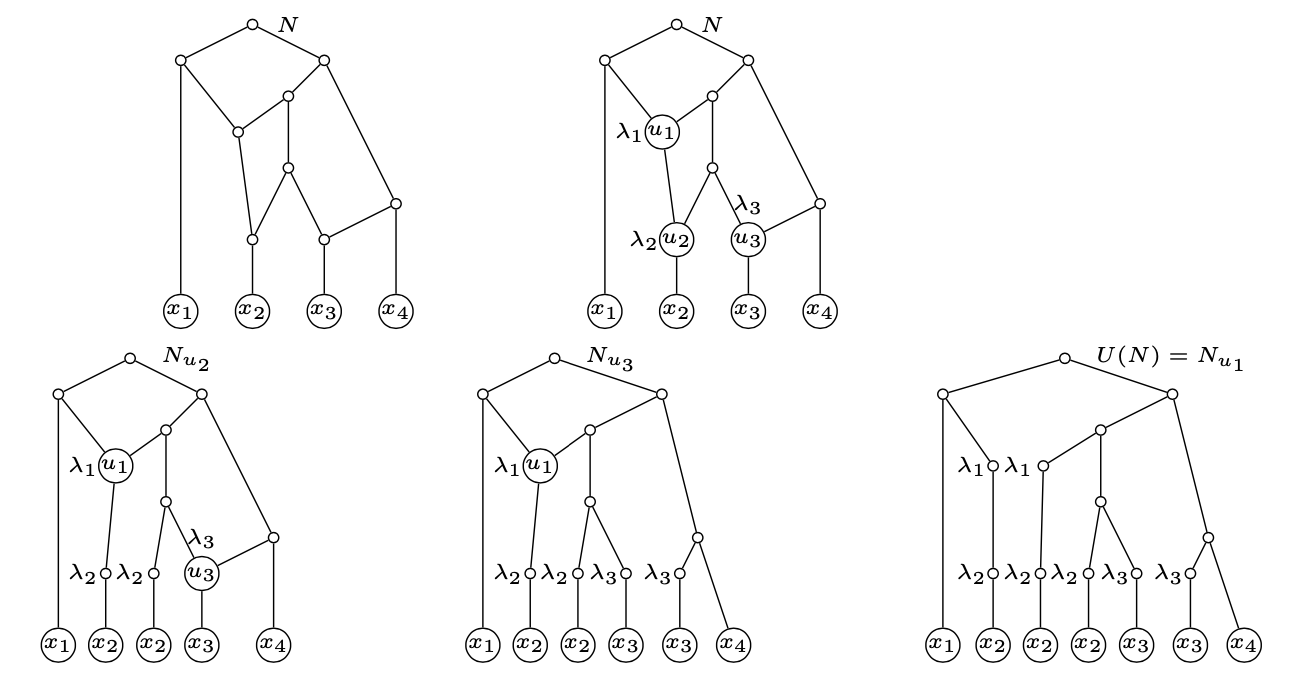}
    \caption{{\bf The unfolding of  an IMLN}. {Top two figures: A phylogenetic network $N$ on $\{x_1,x_2,x_3,x_4\}$, and the IMLN obtained by considering the labelling function over $R(N)$ given by $\ell(u_i)=\lambda_i$ for $i\in\{1,2,3\}$. Notice that $N$ is an internally labelled phylogenetic network. The three figures below are the sequence of unfoldings $(N_{u_2}, N_{u_3}, N_{u_1})$ associated to the compatible sequence of reticulations $(u_2,u_3,u_1)$. Following  the introduced terminology, $N_{u_2}=U(N,u_2)$, $N_{u_3}=U(U(N,u_2),u_3)$ and $N_{u_1}=U(U(U(N,u_2),u_3),u_1)=U(N)$. Note that $u_2,u_3 \in R_{\min}(N)$ and $u_1 \notin R_{\min}(N)$, there is a path from $u_1$ to $u_2$ in $N$.}}
    \label{fig:unf}
\end{figure}

Therefore, given a compatible sequence $(u_1, u_2, \ldots, u_r)$ of all the elements of $R(N)$,  and its associated sequence $(N, N_{u_1}, N_{u_2},\ldots, N_{u_r})$,  we define the \textit{unfolding} of an {IMLN} $N$, denoted by $U(N)$, by means of the equation $U(N) = N_{u_r}$. We may refer to such a sequence as a \textit{sequence of unfoldings}.
See Fig \ref{fig:unf} for an example of a sequence of unfoldings for an IMLN; in fact for an internally labelled phylogenetic network.

Now, we are interested in the ``reverse'' process to unfolding. Roughly speaking, we are interested in formally defining a way to ``fold'' an IMLT to recover the IMLN from which it comes. 
We can, given an IMLN $N$, also define a partial order over the set of elementary nodes $E(N)$ by saying that for any two $u,v\in E(N)$, $u\leq_E v$ if and only if there exist $u', v'\in E(N)$ with $\ell(u) = \ell(u')$ and $\ell(v) = \ell(v')$ and a directed path from $v$ to $u$. We call the set of elementary nodes that are maximal under this order $E_{\max}(N)$.

\begin{lem} \label{lem:elem}
Let $(N, N_{u_1}, N_{u_2},\ldots, N_{u_r})$ be a sequence of unfoldings of an internally labelled phylogenetic network $N$. For any $N_{u_i}$ in it and for every $u\in E_{\max}(N_{u_i})$, there exists exactly another $v\in E_{\max}(N_{u_i})$ such that $\ell(u) = \ell(v)$ and the IMLT's $N_{u_i}(u)$ and $N_{u_i}(v)$ are isomorphic.  
\end{lem}
\begin{pf}
Let $N_{u_i}$ be one of the IMLN's in the sequence of unfoldings. Let $N'=N_{u_{i-1}}$, with $N'=N$ when $i=1$.  By construction, $u_{i}\in R_{\min}(N')$. 
\\
Since $N_{u_i}=U(N',u_i)$, the IMLT $N'(u_i)$  is duplicated; say $u$ and $v$ the two resulting copies of $u_i$ in $N_{u_i}$, we have $\ell(u) = \ell(v)$ and $N_{u_i}(u)=N_{u_i}(v)$. Moreover, $u, v \in E_{\max}(N_{u_i})$; otherwise, if $u$ (or $v$) is not maximal under the order $\leq_E$ in $N_{u_i}$, it means that there are  $w, w'\in E(N_{u_i})$ with $\ell(w)=\ell(w')$ such that there is a path from $w$ to $u$. By Remark \ref{rem:unfolded-paths} this path is preserved in every $N_{u_j}$ with $j<i$. Since the labelling function $\ell$ is injective over reticulation nodes and $N$ has not elementary nodes, this means that the pair $w, w'$ corresponds to a reticulation node in some $N_{u_j}$ with $j<i$; equivalently, this is a reticulation node equal to some $u_j$ with $j<i$. This leads to a contradiction with the fact that the sequence $(u_1,u_2,\ldots, u_r)$ is compatible.
If we consider a maximal element in $N_{u_i}$ different to the two coming from the duplication of $u_i$ in $N'$, the previous argument can be reproduced similarly. These pair of maximal elements are preserved as maximal in every $N_{u_j}$ with $j<i$ right up until the unfolding on this reticulation is produced. This proves that the IMLT's rooted on the corresponding copies of it are also preserved until $N_{u_i}$ is reached.  
\end{pf}

In particular, in the proof of Lemma \ref{lem:elem}, and following the same notation, we show that the node $u_i$ is maximal under the $\leq_E$ order in $N_{u_i}$. Notice also that this could be false if elementary nodes are allowed in the initial IMLN $N$.

\begin{prop}\label{lem:elem-order}
Let $(N, N_{u_1}, N_{u_2},\ldots, N_{u_k})$ be a sequence of unfoldings of an internally labelled phylogenetic network $N$. For any $N_{u_i}$ in it, let $w\in E_{\max}(N_{u_i})$. Then, $v\in E(N_{u_i})$ is such that $v\leq_E w$ if and only if $v\leq_R w$ in $R(N)$.
\end{prop}
\begin{pf}

We begin by the ``if'' direction. If $v,w$ are such that $v\leq_R w$  when seen as reticulation nodes in $N$, there exists at least a path from $w$ to $v$. Now, since $w\in E_{\max}(N_{u_i})$, by Lemma \ref{lem:elem}, there exists $w'\in E_{\max}(N_{u_i})$  such that $\ell(w) = \ell(w')$ and $N_{u_i}(w)=N_{u_i}(w')$, \textit{via} an isomorphism $f$. Then, since by hypothesis $v\in E(N_{u_i})$ and, by Remark \ref{rem:unfolded-paths}, the path from $w$ to $v$ in $N$ is preserved in $N_{u_i}$, there exist paths from $w$ to $v$ and from $w'$ to $f(v)$ in $N_{u_i}$, such that $\ell(v)=\ell(f(v))$ and therefore $v\leq_E w$ in $N_{u_i}$.

On the opposite direction, suppose that $v,w$ are such that $v\leq_E w$. Again by Lemma \ref{lem:elem}, in $N_{u_i}$ there exists $w'$ such that $\ell(w) = \ell(w')$ and $N_{u_i}(w)=N_{u_i}(w')$ \textit{via} an isomorphism $f$. Since $v\leq_E w$, there exists a path from $w$ to $v$ and  a path from $w'$ to $f(v)$ and $\ell(v) = \ell(f(v))$. Now, since there are no elementary nodes in $N$, there must exist $j<i$ such that in $N_{u_j}$ (it could be that $N_{u_j}=N$), the nodes $v$ and $w$ are reticulations. By Remark \ref{rem:unfolded-paths}, this implies that there would exist a path from $w$ to $v$ in $N_{u_j}$, and therefore $v\leq_R w$ in $N_{u_j}$, and so in $N$. Thus concludes the proof.
\end{pf}

Given $N$ an IMLN, $u\in R_{\min}(N)$ and $U(N, u)$, we would like to consider $N$ to be the result of a folding operation over $U(N, u)$: $N = F(U(N,u), u)$, for some suitable $F$. For any unfolding sequence $(N, N_{u_1}, N_{u_2},\ldots, N_{u_r})$, we say that each of its members is a \emph{(phylogenetic) pseudo-network} ---in particular, they are IMLN's. Equivalently, we can define a pseudo-network recursively as follows: let $N$ be an IMLN; it is a pseudo-network if it satisfies the following three conditions:

\begin{itemize}
    \item[(i)] no reticulation node descends from an elementary node; 
    \item[(ii)] for any $u\in E_{\max}(N)$ there exists $v\in E_{\max}(N)$ such that $\ell(u) = \ell(v)$ and $N(u) = N(v)$ as IMLT's; 
    \item[(iii)] for any $u\in E_{\max}(N)$, the IMLN obtained by the process of
    \begin{enumerate}
        \item considering the node $v\in E_{\max}(N)$ such that $\ell(v) = \ell(u)$ and $N(u) = N(v)$, and the parent of $v$, say $v^{(1)}$;
        \item deleting $N(v)$, as well as the edge $(v^{(1)},v)$;
         \item adding the arc $(v^{(1)},u)$,
\end{enumerate} 
is also a pseudo-network.
\end{itemize}

The IMLN obtained by the process described in (iii) is denoted by $F(N,u)$, and called the \emph{folded IMLN of $N$ at $u$}. Notice that if $u,v\in E_{\max}(N)$ are such that $\ell(u) = \ell(v)$, then $F(N, u) = F(N, v)$.

\begin{lem}\label{lem:folding}
Let $N$ be a pseudo-network and $u\in R_{\min}(N)$. Then, 
$$
F(U(N, u), u) = N.
$$
\end{lem}
\begin{pf}
Let $N'=U(N, u)$. Since $u\in R_{\min}(N)$, then $N(u)$ (the tree rooted at $u$) is an IMLT. Let $v_1, v_2$ be the parents of $u$ in $N$. When $N(u)$ is duplicated in the unfolding process, $u$ and a new copy of it, say $v$, are  elementary nodes and the roots of $N'(u)$ and $N'(v)$ respectively, such that $N'(u)=N'(v)$. Moreover, $(v_1,u), (v_2,v)$ are arcs in $N'$. Since $u \in E_{\max}(N')$ (because $u\in R_{\min}(N)$), by Lemma \ref{lem:elem}, $v$ is the other node in $E_{\max}(N')$,  such that $\ell(u) = \ell(v)$ and $N'(u)=N'(v)$. By definition of the folding process of $N'$ at $u$, the IMLT $N'(v)$ and also the arc $(v_2,v)$ are deleted and a new arc $(v_2,u)$ is created. This results in a reticulation node $u$ with parents $v_1$ and $v_2$ which is the root of $N'(u)$. Since $N(u)=N'(u)$, then  $F(N', u) = N$.
\end{pf}

Given $N$ an IMLN and $(N, N_{u_1}, N_{u_2}, \ldots, N_{u_r})$ a sequence of unfoldings, by Lemma \ref{lem:folding} we have that $N_{u_i} = F(N_{u_{i+1}}, u_{i+1})$ and that $N = F(N_{u_1}, u_1)$. Therefore, we derive the following result. 

\begin{cor}\label{cor:folding}
Let $N$ be an internally labelled phylogenetic network and  $(N, N_{u_1}, N_{u_2}, \ldots, N_{u_r})$ any sequence of unfoldings. Then
$$
N = F(F(F(\ldots F(U(N), u_r)\ldots), u_2) u_1).
$$
\end{cor}

Note that, similarly as we have done by the equivalent compatible sequences, there is not a unique way to recover the IMLN $N$ by applying a set of foldings.

If $N$ is a pseudo-network we know that it is the product of a sequence of unfoldings performed over an {IMLN}, $N'$. We can then rewrite Corollary \ref{cor:folding}, by defining a function $F$ from the set of pseudo-networks to the set of {IMLN's} by $F(N):=N'$. Hence,

\begin{cor}\label{cor:fu}
Let $N$ be an internally labelled phylogenetic network. Then
$$
N = F(U(N)).
$$
\end{cor}

This result is the analogue of the concept of stable networks in Section 4 of~\cite{huber2016folding}.  The key difference here is that we allow elementary nodes.

\section{A polynomial for internally multi-labelled phylogenetic networks}\label{sec:polynomial}

Given a phylogenetic network $N$ on $X$, 
one can obtain a rooted tree by removing one incident arc to each reticulation node. These (sub)trees could contain elementary  nodes, and its leaves might be labelled in $X$ (the leaves from $N$) and other sets different from it (for instance when the single outgoing arc to a reticulation is removed). Those trees become unrooted if the direction of the arcs is suppressed (particularly, the root becomes a degree two node) and are called \emph{embedded spanning trees} if its set of leaves is exactly $X$.
Tree-child phylogenetic networks are characterized by their set of embedded spanning trees \cite{francis2018identifiability}, 
but not general phylogenetic networks. 

In \cite{janssen2021comparing}, the {Liu} polynomial is generalized to phylogenetic networks by their sets of embedded spanning trees. Roughly speaking, the polynomial of the network is the product of the polynomials of the embedded spanning trees (considering trees with multiplicity). Consequently, this extension is a complete invariant for tree-child networks.

\begin{figure}[!ht]
\centering
 \includegraphics[scale=0.5]{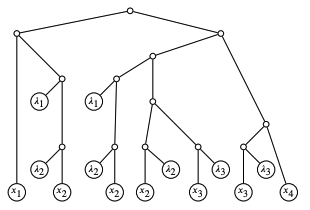}
    \caption{{\bf A multi-labelled tree derived from an internally labelled phylogenetic network}. Let $N$ be the network depicted in Fig \ref{fig:unf}. This figure depicts a decomposition of $N$ resulting in a multi-labelled tree.}
    \label{fig:cesc&new}
\end{figure}

There are some natural extensions of the Liu polynomial to 
 IMLN's that come to mind. The first one, for internally labelled phylogenetic networks, is to completely unfold such a network and, from any elementary node $u$ labelled $\lambda_i$, for some $i\in\{1,\ldots, r\}$ and labels $\lambda_i$ distinguishable from labels $x_i$, grow an arc to a new node $v$, label $v$ as $\lambda_i$, and finally forget the labelling of $u$. Thus, the unfolded IMLT becomes a multi-labelled tree over leaves $\{x_1,\ldots,x_n, \lambda_1, \ldots, \lambda_r\}$. See an example of that decomposition in Fig \ref{fig:cesc&new} from the internally labelled phylogenetic network $N$ depicted in Fig \ref{fig:unf}. By means of Corollary 3.5 in \cite{liu2021tree}, this extension of the polynomial is immediately seen to uniquely characterize an internally labelled phylogenetic network. 

We will here deal with a natural extension that reflects the reticulation process in the sheer morphology of the polynomial, rather than in the name of the variables.

Let $N$ be an IMLN. Then, consider $$p: V(N)\to\ZZ[x_1,\ldots, x_n,\lambda_1,\ldots,\lambda_r, y]$$ to be defined recursively as follows. Let $u\in V(N)$, then:
\begin{itemize}
    \item if $u$ is a leaf, $p(u) = \varphi(u)$;
    \item if $u$ is an internal tree node whose two children are $v_1, v_2$, $ p(u) = y +  p(v_1)  p(v_2)$; 
    \item  otherwise, i.e. if $u$ has only one child $v$ and its associated label is $\lambda_i = \ell(u)$, then $p(u) = \lambda_i p(v)$.
\end{itemize}
Then, let $\rho_N$ be the root of $N$; we define $p(N)$ to be $p(\rho_N)$. Notice that this definition of the polynomial $p$ is given over generic IMLN's.

For example, the polynomial associated to the IMLN represented in Fig \ref{fig:unf} is 
\begin{equation*}
\begin{split}
p(N) = & y+ y^2 + y^3 +  \lambda_1\lambda_2x_2y^3 +  \lambda_3x_3x_4y^2  + \lambda_1\lambda_2x_1x_2y + \lambda_1\lambda_2x_1x_2y^2 + \\ & + \lambda_1\lambda_2\lambda_3x_2x_3x_4y^2 +  \lambda_1\lambda_2^2\lambda_3x_2^2x_3y^2 + \lambda_1\lambda_2^2\lambda_3^2x_2^2x_3^2x_4y + \\ & + \lambda_1\lambda_2\lambda_3x_1x_2x_3x_4y + \lambda_1^2\lambda_2^2x_1x_2^2y^2  + 
\lambda_1^2\lambda_2^2\lambda_3x_1x_2^2x_3x_4y + \lambda_1^2\lambda_2^3\lambda_3x_1x_2^3x_3y + \\ & + \lambda_1^2\lambda_2^3\lambda_3^2x_1x_2^3x_3^2x_4.
\end{split}
\end{equation*}

\begin{prop}\label{prop:irr}
Let $N$ be an IMLN. Then, for any $u\in V(N)$, $p(u)\in\ZZ[x_1,\ldots, x_n,\lambda_1,\ldots,\lambda_r, y]$ is an irreducible polynomial if and only if $u$ is a tree node.
\end{prop}
\begin{pf}
If $u$ is not a tree node the polynomial will not be irreducible, since then there would exist $v\in V(N)$ as the only descendant of $u$, and $p(u) = \ell(u)p(v)$.

It then remains only to see that if $u$ is a tree node, $p(u)$ is irreducible. In this case, either $u$ is a leaf and then $p(u) = \varphi(u) =x_i$ for some $i\in\{1,\ldots, n\}$ and so irreducible, or $u$ has two children and $p(u) = y + \Lambda p(w_1)p(w_2)$, where $\Lambda$ is a product of $\lambda_i$ from $\lambda_1,\ldots,\lambda_r$, and $w_1,w_2$ are the first descendants from $u$ at each side that are tree nodes (they are possibly equal). Now consider the polynomial $p'(u)$ obtained from $p(u)$ by changing every variable $x_1,\ldots, x_n, \lambda_1,\ldots, \lambda_r$ for, say, $x_1$.  Then, it can be seen that $p'(u)$ satisfies Eisenstein's irreducibility criterion in $\ZZ[y][x_1]$ (which is an unique factorization domain, UFD) applied to the ideal $\langle y\rangle$, and so $p(u)$ is irreducible when seen as a polynomial in $\ZZ[y][x_1,\ldots,x_n,\lambda_1,\ldots,\lambda_r]$. But, since $y$ does not divide $p(u)$, then $p(u)$ is also irreducible in $\ZZ[x_1,\ldots,x_n,\lambda_1,\ldots,\lambda_r, y]$.
\end{pf}

The next proposition will show that the polynomial is conserved throughout a sequence of unfoldings, and therefore will allow us to compute it over any of its members without distinction. In particular, it can be computed on the unfolding of the network. 

\begin{prop}\label{prop:pol-fold}
Let $N$ be an IMLN, and $(N, N_{u_1}, N_{u_2},\ldots, N_{u_r})$ be a sequence of unfoldings. Then, $p(N) = p(N_{u_1})$ and, for any $i\in\{1,\ldots, r-1\}$, $p(N_{u_{i+1}}) = p(N_{u_{i}})$.
\end{prop}
\begin{pf}
Let $N'$ be an IMLN, and $u\in R_{\min}(N')$. If we are able to show that $p(N') = p(U(N', u))$, then the proposition will hold.
Let $v^{(1)}, v^{(2)}$ be the parents of $u$, in $U(N', u)$ each of them will be the parent of at least one elementary node $u_x$, $x\in\{1,2\}$, which will be the root of a copy of the IMLT $N'(u)$, and by construction $p(u_1) = p(u_2) = p(u) = p(N'(u))$. Now, by the definition of the polynomial, $p(v^{(x)})$ will be the same in $N'$ and in $U(N', u)$. Therefore, $p(N') = p(U(N', u))$.
\end{pf}

We now introduce two remarks, the first concerning the interpretation of the coefficients and, the second, about the reconstruction of the unfolding of an IMLN from the polynomial if it characterizes the IMLN.

\begin{figure}[!ht]
    \centering
 \includegraphics[scale=0.5]{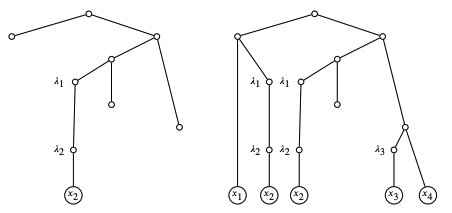}  
    \caption{{\bf Two primary subtrees of $U(N)$}. {Let $N$ be the IMLN depicted in Fig \ref{fig:unf}. The figure depicts two primary subtrees of $U(N)$ corresponding to the terms $\lambda_1\lambda_2x_2y^3$ (left), and $\lambda_1^2\lambda_2^2\lambda_3x_1x_2^2x_3x_4y$ (right), of the polynomial $p(N)$.}}
    \label{fig:coefs}
\end{figure}

\begin{rmk}
The interpretation of the coefficients of the polynomial $p(N)$ can be extended from Lemma 2.4 in \cite{liu2020polynomial} by slightly modifying the definition of primary subtrees to the IMLT $T=U(N)$.  
Let a \emph{primary subtree} $S$ of $T$ be a rooted subtree of $T$ such that $S$ shares the same root node with $T$ and any leaf node in $T$ is either a leaf node in $S$ or a descendant of a leaf node in $S$ which does not come from an elementary node. 

Then, if we represent $p(N)$ as $$\sum c(\gamma_1, \ldots, \gamma_r, \alpha_1, \ldots, \alpha_n, \beta) \lambda_1^{\gamma_1}\cdots \lambda_r^{\gamma_r} x_1^{\alpha_1} \cdots x_n^{\alpha_n}y^{\beta},$$ each one of its coefficients counts the number of primary subtrees of $U(N)$ satisfying that: 
\begin{itemize}
    \item $\gamma_i$ (for $i\in\{1, \ldots, r\}$) is the number of nodes labelled by $\lambda_i$ of these subtrees;
    \item $\alpha_i$ (for $i\in\{1, \ldots, n\}$) is the number of leaf nodes labelled by $x_i$ of these subtrees which are also leaves in $U(N)$; 
    \item $\beta$ is the number of leaf nodes of these subtrees which are internal nodes in $U(N)$.
\end{itemize}
 See Fig \ref{fig:coefs} for the interpretation of some of the terms of the polynomial $p(N)$ of the IMLN $N$ depicted in Fig \ref{fig:unf}. Notice that these primary subtrees can then be folded into a sort of ``sub-primary networks''.
 \end{rmk}
 
\begin{rmk} 
In this remark we shall give a first approximation to the problem of reconstructing the Newick string of an IMLT $U(N)$ from $p(N)$, in the case where the polynomial characterizes $N$. Roughly speaking, we proceed as follows: start by substracting $y$ from $p(N)$ and then factor $p(N)-y=q_1 \cdot q_2$. Then the Newick string to consider is $(q_1,q_2)$. From now on, whenever it is possible to substract $y$ from a polynomial $q$, do so. If the factorization involves only two members, $q=q_1\cdot q_2$, then proceed as before and replace $q$ by $(q_1,q_2)$. Otherwise, there could be conflicts in terms of deciding how to group members in a factorization of type $$\prod_{j \in J\subseteq\{1,\ldots,r\}} \lambda_j \prod_k q_k,$$ where $q_k$ are polynomials.  But there will always be in the queue of factorizations pending to be grouped, a pair of them where a ``minimum'' monomial of type $\lambda_i \cdot q_s$  is common in both; this allows one to determine that there is an arc from an elementary node labelled by $\lambda_i$ to the subtree determined by the polynomial $q_s$. In terms of the Newick string, it could be replaced by $(\lambda_i(q_s))$. 
\end{rmk}

\begin{figure}[!ht]
    \centering
 \includegraphics[scale=1.1]{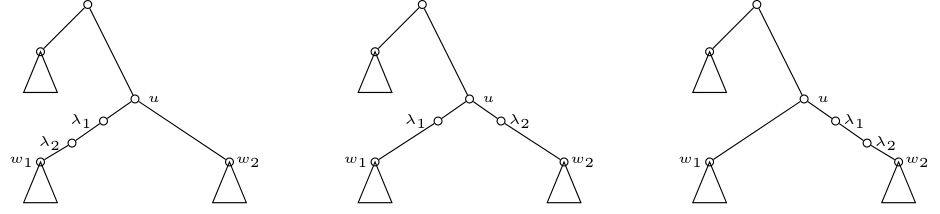}  
    \caption{{\bf Non-isomorphic IMLT's}. {Three non-isomorphic IMLT's presenting the same polynomial at $u$.}}
    \label{fig:strongpaths}
\end{figure}

We are now specially interested in determining under which conditions the polynomial associated to an IMLN uniquely characterizes it. Note that this is not always the case, indeed for IMLT's. See for instance the three representations of IMLT's in Fig \ref{fig:strongpaths}. The polynomial fails to correctly distinguish between them. Roughly speaking, looking at the polynomials of the elementary vertices we could readily distinguish between the three possibilities, but we cannot do so by only looking at $p(u)$, since $p(u)=y+\lambda_1\lambda_2p(w_1)p(w_2)$.

\subsection{Strong paths}

 We shall now present a series of definitions. Let $N$ be an IMLN, and $u,v\in V(N)$. If there exists a path from $u$ to $v$ consisting only of elementary or reticulation nodes, we say that $u$ is a \textit{strong ancestor} of $v$, and that $v$ is a \textit{strong descendant} of $u$. Such a path is called a \textit{strong path}. For example, by considering the situation in Fig \ref{fig:strongpaths}, we can see that in all three cases $w_1, w_2$ strongly descend from $u$.

\begin{lem}\label{lem:ret-unique}
Let $N$ be an internally labelled phylogenetic network, and $v_1,v_2$ two reticulation nodes. If $p(v_1) = p(v_2)$, then $v_1 = v_2$.
\end{lem}
\begin{pf}
Let $w_1$ be the child of $v_1$; by the definition of the polynomial, $p(v_1) / p(w_1) = \lambda_i$ for some $\lambda_i\in\{\lambda_1,\ldots,\lambda_r\}$. Since $p(v_1) = p(v_2)$, it also means that $p(v_2) / p(w_1) = \lambda_i$, but since $N$ is an internally labelled phylogenetic network this implies that $v_2$ is a parent of $w_1$ and that $\ell(v_2) = \lambda_i$. Thus, they are the same node.
\end{pf}

\begin{lem}\label{lem:ret-desc}
Let $N$ be an internally labelled phylogenetic network, and $v$ a reticulation node in it. A node $u$ is a strong ancestor of $v$ if, and only if, one of the two following conditions happens:
\begin{itemize}
    \item $p(v)\mid p(u)$, and then $u$ is a reticulation node, or
    \item $p(v)\mid (p(u) - y)$, and then $u$ is a tree node.
\end{itemize}
\end{lem}
\begin{pf}
By the definition of the polynomial and Lemma \ref{lem:ret-unique}.
\end{pf}

Now, if we want to compare two IMLN's on the same sets of labels $\{x_1, \ldots, x_n\}$ and $\{\lambda_1, \ldots, \lambda_r\}$, we should take into account the possibility that two of them are isomorphic up to a permutation of the labels. In order to express this possibility, let $\sigma:\{x_1, \ldots, x_n, \lambda_1, \ldots, \lambda_r\}\to \{x_1, \ldots, x_n, \lambda_1, \ldots, \lambda_r\}$ be a permutation such that $\sigma(X) = X$ (i.e., that fixes the sets of labels of the leaves and of the elementary or reticulation nodes). Given an IMLN $N$, we denote by $^\sigma N$ the network isomorphic to $N$ that has all its labels permuted according to $\sigma$, and by $^\sigma p(N)$ we mean $p(^\sigma N)$ or, equivalently, the polynomial that has all its variables changed according to $\sigma$.

\begin{defn}\label{def:strong-path}
Let $N_1, N_2$ be two IMLN's, and $\sigma$ a permutation of their labels such that $\sigma(X) = X$. We say that $N_1$ and $N_2$ are \textit{equivalent} modulo \textit{strong paths} if the following three conditions are satisfied:
\begin{enumerate}
    \item[(i)] $p(N_1) =\ ^\sigma p(N_2)$;
    \item[(ii)] there exists a bijection $f$ between the sets of tree nodes of $N_1$ and $N_2$ such that, if $u, v$ are tree nodes and $v$ is a strong descendant of $u$, then $f(v)$ is a strong descendant of $f(u)$;
    \item[(iii)] for any tree node $u$ in $N_1$, $p(u) =\ ^\sigma p(f(u))$.
\end{enumerate}
\end{defn}

Being equivalent \textit{modulo} strong paths is an equivalence relation.

\begin{rmk}\label{rem:strong-paths}
The above definition can also be easily stated exclusively in terms of strong paths, which are intrinsic to the IMLN. However, the definition in terms of the polynomial is more tractable and concise.
\end{rmk}

Notice that all the IMLT's in Fig \ref{fig:strongpaths} are equivalent \textit{modulo} strong paths. Indeed, we present the following theorem:

\begin{thm}\label{thm:imlt-inj}
 Let $N_1, N_2$ be two IMLN's, and $\sigma$ a permutation of their labels such that $\sigma(X) = X$. Then, $p(N_1) =\ ^\sigma p(N_2)$ if, and only if, $N_1$ and $N_2$ are equivalent \emph{modulo} strong paths.
\end{thm}
\begin{pf}
The ``if'' part of the implication is direct by the first condition of the definition of equivalence \textit{modulo} strong paths.

Suppose now that $p(N_1) =\ ^\sigma p(N_2)$, and let us show that $N_1$ and $N_2$ must be equivalent. We first see that there exists a bijection $f$ between the sets of tree nodes of $N_1$ and $N_2$ such that for any tree node $u$ in $N_1$, $p(u) =\ ^\sigma p(f(u))$. We will use the following inductive schema: we shall prove that, if $u$ is a tree node in $N_1$ and $f(u_1)$ is a tree node in $N_2$ such that $p(u) =\ ^\sigma p(f(u))$, then if $w_1, w_2$ in $N_1$ are the two tree nodes that strongly descend from $u_1$, then the two tree nodes $w_1', w_2'$ that strongly descend from $f(u)$ in $N_2$ are such that $p(w_1) =\ ^\sigma p(w_1')$ and $p(w_2) =\ ^\sigma p(w_2')$. Then, we will provide tree nodes $u_1, u_2$ in $N_1$ and $N_2$, respectively, from which all other tree nodes will descend and such that $p(u_1) =\ ^\sigma p(u_2)$.

Let $u$ be a tree node in $N_1$, and $w_1,w_2$ be the two tree nodes that strongly descend from it. Then, $p(u) = y + \mu_1\cdot\ldots\cdot\mu_{r'}p(w_1)p(w_2)$, for $\mu_1,\ldots,\mu_{r'}\in\{\lambda_1,\ldots,\lambda_r\}$. Then, if $p(u) =\ ^\sigma p(f(u))$, $\mu_1\cdot\ldots\cdot\mu_{r'}p(w_1)p(w_2) =\ ^\sigma\mu_1'\cdot\ldots\cdot\ ^\sigma\mu_{r'}'\ ^\sigma p(w_1')^\sigma p(w_2')$, where $w_1', w_2'$ are the tree nodes that strongly descend from $f(u)$ in $N_2$; but since $p(w_1), p(w_2)$ are both irreducible and different from any $\lambda_i$, then it must happen that (without loss of generality) $p(w_1) =\ ^\sigma p(w_1')$ and $p(w_2) =\ ^\sigma p(w_2')$. Thus, set $f(w_1) = w_1'$ and $f(w_2) = w_2'$.

We will now show that there is a tree node in both $N_1$ and $N_2$ such that any other tree node descends from it. Suppose that the root of $N_1$, say $\rho_1$, is a tree node; if so, since $p(N_1) =\ ^\sigma p(N_2)$ and by Proposition \ref{prop:irr}, the root of $N_2$, say $\rho_2$, must also be a tree node. Therefore, any other tree node in their respective IMLN's must descend from them, and furthermore $p(\rho_1) =\ ^\sigma p(\rho_2)$. Set $f(\rho_1) = \rho_2$.

Finally, suppose that $\rho_1$ is not a tree node; then, $p(\rho_1)$ is not an irreducible polynomial, and therefore neither will $^\sigma p(\rho_2)$. Let $w_1$ be the only tree node strongly descending from $\rho_1$ in $N_1$. It is straightforward to see that, if $w_1'$ is the only tree node strongly descending from $\rho_2$ in $N_2$, then $p(w_1) =\ ^\sigma p(w_1')$. In both cases, any other tree node in the network will descend from them. Therefore, set $f(w_1) = w_1'$.
\end{pf}

Now, the question arises: under which conditions can we say that two internally labelled phylogenetic networks that are equivalent \textit{modulo} strong paths are actually isomorphic?

\subsection{Separability: a sufficient condition}

In this part we shall give a sufficient condition for two internally labelled phylogenetic networks to be completely characterized by the  polynomial. In order to do so, we will work with the immediate neighbourhood of any tree node.

\begin{figure}[!ht]
    \centering
 \includegraphics[scale=1.5]{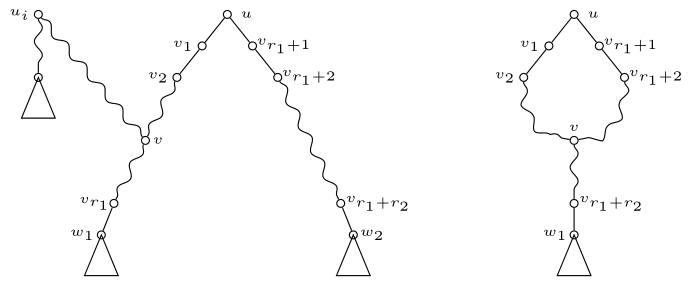}  
    \caption{{\bf Strong paths from a tree node}. {A tree node $u$ and its strong descendants $w_1$ and $w_2$ (left) or $w_1$ (right). The curly paths represent strong paths. The nodes $v$ and $u_i$ are used in the proof of Lemma \ref{lem:entry}}.}
    \label{fig:lem_separ}
\end{figure}

Let $N$ be a phylogenetic network, and let $u$ be a tree node in $N$. Let $w_1, w_2$ be the two (possibly equal) tree nodes that strongly descend from it. Let $v_1,\ldots, v_{r_1},\ldots, v_{r_1+r_2}$ be the reticulation nodes in the strong paths from $u$ to $w_1$ and $w_2$, and suppose that there are $r_1$ such nodes in the path from $u$ to $w_1$ and $r_2$ in the other. See Fig \ref{fig:lem_separ}.
Let $U(u) = \{u_1,\ldots, u_k\}$ be the set of all the tree nodes that are strong ancestors of $w_1$ or $w_2$ different from $u$. Note that the node $u_i$ in Fig \ref{fig:lem_separ} (left) is a node in $U(u)$. In what follows, we will allow ourselves to write $U$ if the context is sufficiently clear. We will present now the following lemma.

\begin{lem}\label{lem:entry}
Consider the situation above. Let $v$ be a reticulation node from the collection $v_1,\ldots, v_{r_1+r_2}$. Then, there are two possibilities: 
\begin{itemize}
    \item both its parents are nodes from $v_1,\ldots, v_{r_1+r_2}$, or
    \item there exists at least one tree node $u_i\in U$ such that there is a strong path from $u_i$ to $v$ not containing any other reticulation node $v_1,\ldots, v_{r_1+r_2}$.
\end{itemize}
Furthermore, the first possibility can only happen for \emph{one} reticulation node in $v_1,\ldots, v_{r_1+r_2}$, and it will hold if, and only if, $w_1=w_2$.
\end{lem}
\begin{pf}
Suppose that $v$ is the first reticulation node (counting by proximity to $u$) that satisfies the first condition (this makes sense, since our networks are binary). In this situation, from it emerges only one path up to the next tree node. But since $N$ is binary, the two paths that emerged from $u$ are now confounded in the only path from $v$ to the next tree node, $w_1 = w_2$. See Fig \ref{fig:lem_separ}, right.  Therefore, since there is now only one path of reticulation nodes, no other node in it can satisfy the first condition.

If $v$ does not satisfy the first condition, one of its parents must not be from $v_1,\ldots, v_{r_1+r_2}$. Let $u_i$ be a tree node strong ancestor of such a parent of $v$. The pair $v, u_i$ satisfies the second condition. See Fig \ref{fig:lem_separ}, left.
\end{pf}

We say that a tree node $u_i\in U(u)$ \textit{enters the neighbourhood of $u$ at $v$} if the pair $v,u_i$ satisfies the second condition of Lemma \ref{lem:entry}. If the context is sufficiently clear, we shall only say that it \textit{enters at $v$}. Likewise, we say that $v$ is the \textit{entry of $u_i$ to the neighbourhood of $u$} (or that it is just its \textit{entry}).

We can then divide the set $U$ into five sets: let $v^{(x)}$, $x\in\{1,2\}$, be the two children of $u$, then we define
    \begin{align*}
        U_1^{(x)} = &\ \{u_i\in U : u_i \textrm{ enters the neighbourhood of } u \textrm{ at only one} \\
        & \qquad\textrm{ reticulation node } v \textrm{ that is a strong descendant of } v^{(x)}\}, \\
        U_2^{(x)} = &\ \{u_i\in U : u_i \textrm{ enters the neighbourhood of } u \textrm{ at two (possibly equal)} \\
        & \qquad\textrm{ reticulation nodes } v_1, v_2 \textrm{ that are strong descendants of } v^{(x)}\}, \\
        U_3 = &\ U\setminus(U_1^{(1)}\cup U_1^{(2)} \cup U_2^{(1)} \cup U_2^{(2)}).
    \end{align*}
Notice that, if $w_1 \neq w_2$, then 
    \begin{align*}
        U_3 =\{u_i\in U : u_i \textrm{ is a strong ancestor of both } w_1 \textrm{ and } w_2\}. 
    \end{align*}

\begin{figure}[!ht]
    \centering
 \includegraphics[scale=1.5]{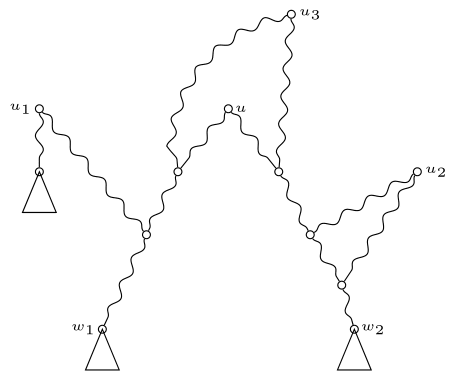}  
    \caption{{\bf Division of $U(u)$}. {Three trees nodes evidencing the type of sets in the division of $U(u)$. In this case, $u_1 \in U_1^{(1)}$, $u_2 \in U_2^{(2)}$ and $u_3 \in U_3$.}}
    \label{fig:partitionU}
\end{figure}

The above division $\{U_1^{(1)}, U_1^{(2)}, U_2^{(1)}, U_2^{(2)}, U_3\}$ is a partition of $U$. In Fig \ref{fig:partitionU} three tree nodes $u_1$, $u_2$ and $u_3$ from the set $U=U(u)$ are represented. Note that $u_1 \in U_1^{(1)}$, $u_2 \in U_2^{(2)}$ and $u_3 \in U_3$.   

In general, given all the polynomials evaluated at each tree node of $U$, we cannot deduce the exact configuration of the $v_i$'s. Remember, for instance, for the case where $r_1+r_2=2$, the three situations presented in Fig \ref{fig:strongpaths}.  That is, we had no \textit{a priori} information on which $v_i$ were strong ancestors of $w_1$ and which of $w_2$.  This fact motivates the following definition.

\begin{defn}
Let $N$ be a phylogenetic network and $u$ a tree node in it. Let  $v^{(x)}$, $x\in\{1,2\}$, be the two children of $u$. We say that $u$ is \textit{separable} if either  $v^{(1)}$ and $v^{(2)}$ are tree nodes, or if there exists a tree node $u_1$ different from $u$ such that it satisfies one of the following conditions: 
\begin{itemize}
    \item is a strong ancestor of $v^{(1)}$ (or $v^{(2)}$) but not of any other strong descendant of $u$, or
    \item is a strong ancestor of $v^{(1)}$ (or $v^{(2)}$) and of one of its strong descendants.
\end{itemize}
\end{defn}

\begin{rmk}\label{rem:not-sep}
In this case, the negative definition might be more intuitive. Let $u$ be a tree node with $w_1$ and $w_2$ the tree nodes strongly descended from $u$. Then $u$ is \emph{not separable} if none of its two children $v^{(1)}$ and $v^{(2)}$ are tree nodes, and 
\begin{itemize}
    \item if $w_1\neq w_2$, all the strong ancestors of $v^{(1)}, v^{(2)}$ that are not $u$ are in $U_3(u)$, or
    \item if $w_1 = w_2$ and $v$ is the first reticulation node that is a strong descendant of both $v^{(1)}$ and $v^{(2)}$, then any strong ancestor of $v^{(1)}$ that is not $u$ will be a strong ancestor of a reticulation node in the strong path from $v^{(2)}$ to $v$, and \textit{vice versa}.
\end{itemize}
\end{rmk}

A phylogenetic network is called \textit{separable} if all its tree nodes are so.

\begin{rmk}
Notice that separability is a completely topological condition. Thus, we will use it indistinguishably for phylogenetic networks and internally labelled phylogenetic networks.
\end{rmk}

The key point in separability is that given $u$ a separable tree node and all the polynomials of the tree nodes that are strong ancestors of $w_1$ and $w_2$, we can actually identify the polynomial $p(u_1)$ of the tree node that satisfies the conditions of the definition, and thus we can identify which reticulation nodes descend from $v^{(1)}$ and which from $v^{(2)}$. Indeed: if $w_1\neq w_2$,  $p(u_1)$ will be such that $p(w_1)$ divides $p(u_1)-y$ but $p(w_2)$ does not, and contains the largest number of $\lambda_1,\ldots,\lambda_r$ dividing $p(u)-y$.
If $w_1 = w_2$, the argument is analogous using that  $p(w_1)^2$ is not a divisor of $p(u_1)-y$. 
As a result, we are able to deduce that $p(v^{(x)})= \mu_1^{(x)}\ldots\mu_{r_x}^{(x)}p(w_x)$, $x\in\{1,2\}$, for $\mu_1^{(x)}\ldots\mu_{r_x}^{(x)}$ dividing $p(u)-y$. Thus, we are able to ``separate'' $p(u)$ into the contributions from $p(v^{(1)})$ and $p(v^{(2)})$.

\begin{figure}[!ht]
    \centering
 \includegraphics[scale=1.2]{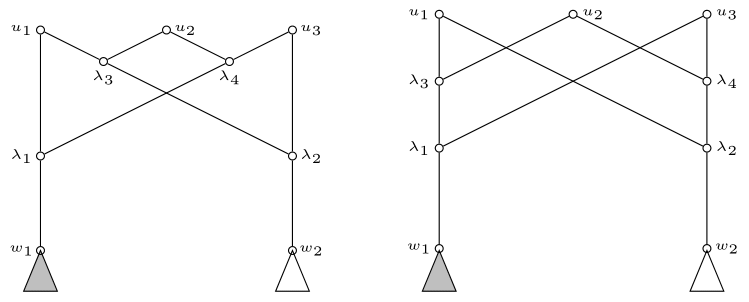}  
    \caption{{\bf Non separable internally labelled phylogenetic networks}. {None of the nodes $u_1,u_2,u_3$ are separable. The filled and non-filled triangles pending from $w_1$ and $w_2$ represent non-isomorphic sub-networks.}}
    \label{fig:non_separ}
\end{figure}

Fig \ref{fig:non_separ} depicts two sub-networks which can be part of internally labelled phylogenetic networks (and then part of the underlying phylogenetic networks) that are not separable.  Notice that they are not separable at any of the nodes $u_1,u_2,u_3$. The filled triangle and non-filled triangle pendant at $w_1$ and $w_2$ represent non-isomorphic sub-networks (for example a leaf and a cherry). Note that in both cases we have the same polynomials at $u_i$, namely $p(u_1)= y+ \lambda_1\lambda_2\lambda_3p(w_1)p(w_2)$, $p(u_2)= y+ \lambda_1\lambda_2\lambda_3\lambda_4p(w_1)p(w_2)$ and $p(u_3)=y+ \lambda_1\lambda_2\lambda_4p(w_1)p(w_2)$. Thus, we can not distinguish between the sub-networks when looking at $p(u_1)$, $p(u_2)$, $p(u_3)$.

\begin{lem}\label{lem:diff-equiv}
Let $N$ be an internally labelled phylogenetic network, and $u_1$ a tree node in it such that it is one of the deepest tree node (i.e., one for
which exists path of maximal length from the root to it) satisfying the following condition: there exists another tree node $u_2$ such that $p(u_1) = p(u_2)$. Then, $u_1$ and $u_2$ must have the same set of children.
\end{lem}
\begin{pf}
If $u_1$ is a leaf, there is nothing to prove, because all the leaves have a different label. Then if $p(u_1) = p(u_2)$, and $p(u_1)=\varphi(u_1)$, we must have $u_2=u_1$. In the other case, let $v^{(1)},v^{(2)}$ be the two children of $u_1$; since $p(v^{(1)})$ and $p(v^{(2)})$ both divide $p(u_2)-y$ and are unique (because $u_1$ is one of the deepest node satisfying the condition in the statement of the lemma), $u_2$ is a strong ancestor to both of them. Therefore, $v^{(1)}, v^{(2)}$ must be reticulation nodes.

We write 
\[p(u_1) = y + \mu_1^{(1)} \cdot \ldots \cdot \mu_{r_1}^{(1)} \mu_1^{(2)} \cdot \ldots \cdot \mu_{r_2}^{(2)} p(w_1) p(w_2),\] 
where $w_1,w_2$ are the tree nodes that strongly descend from $u_1$, $p(v^{(x)}) = \mu_1^{(x)} \cdot \ldots \cdot \mu_{r_x}^{(x)} p(w_x)$ for $x\in\{1,2\}$, and $\mu_{i}^{(1)}, \mu_{j}^{(2)} \in \{\lambda_1, \ldots, \lambda_r\}$. From $v^{(x)}$ to $w_x$ there is only one strong path of length $r_x$, and since $u_2$ is a strong ancestor of both $v^{(1)}$ and $v^{(2)}$ there are $r_1+r_2$ polynomials $\lambda_1,\ldots,\lambda_r$ that divide $p(u_2)-y$. But these are exactly the number  of polynomials in $\lambda_1,\ldots,\lambda_r$ that must divide $p(u_2)-y$, since $p(u_1) = p(u_2)$.
\end{pf}

\begin{lem}\label{lem:sep-diff}
Let $N$ be an internally labelled separable phylogenetic network, and $u_1, u_2$ two internal nodes in it. Then, $p(u_1) = p(u_2)$ if, and only if,  $u_1=u_2$.
\end{lem}
\begin{pf}
The ``if'' part is trivial by the definition of the polynomial. By Lemma \ref{lem:ret-unique}, if either $u_1$, or $u_2$ is a reticulation node, the result is proven. Therefore, assume that $u_1, u_2$ are both tree nodes, and suppose, for the sake of contradiction, that $u_1\neq u_2$. Furthermore, assume that $u_1$ is one of the deepest nodes satisfying that $p(u_1)=p(u_2)$. \\
By Lemma \ref{lem:diff-equiv}, their sets of children are the same. Let $v_1,v_2$ be the two children of $u_1$ and $u_2$. Then $u_1$ and $u_2$ are the only strong ancestors of both $v_1$ and $v_2$. Moreover, $u_2$ is in $U_3(u_1)$.  This means that $u_1$ is not separable and, therefore, neither is $N$.
\end{pf}

\begin{cor}
If $N$ is a separable phylogenetic network, then there is no pair of tree vertices with the same set of children.
\end{cor}

Note that the other direction of the implication in the above Corollary is false. See for instance the (internally labelled) phylogenetic subnetworks depicted in Fig \ref{fig:non_separ}.  These are non separable and they have different set of children for every pair of tree nodes.

\subsection{Isomorphism of internally labelled phylogenetic networks}

In this part we prove the main theorem of this paper. It roughly says that the polynomial is a complete invariant for the class of internally labelled separable phylogenetic networks up to equivalence \textit{modulo} strong paths.

\begin{lem}\label{lem:iso}
Let $N_1, N_2$ be two internally labelled phylogenetic networks such that, for any $u_1, u_2\in N_x$, $x\in\{1,2\}$, $p(u_1) = p(u_2)$ implies that $u_1=u_2$. Suppose that, for any $u,v\in V(N_2)$, $p(u)\neq p(v)$ if $u\neq v$, and let $f:V(N_1)\to V(N_2)$ be a bijection. If there exists a permutation $\sigma$ of their labels with $\sigma(X) = X$ such that $p(u) =\ ^\sigma p (f(u))$ for any $u\in V(N_1)$, then $f$ is an isomorphism of internally labelled phylogenetic networks.
\end{lem}
\begin{pf}
In order to ease the notation, and without loss of generality, let us assume that $\sigma$ is the identity. The fact that $f$ is a bijection is already required in the statement of the Lemma. Then, we must prove that if $(u,v)\in E(N_1)$, then $(f(u),f(v)) \in E(N_2)$ and that $f$ preserves the labels.\\
Suppose that $u$ is a reticulation node; if $(u,v)\in E(N_1)$, then $p(u) = \lambda_i p(v)$ for some $\lambda_i\in\{\lambda_1,\ldots,\lambda_r\}$. Therefore, $p(f(u)) = \lambda_i p(f(v))$ which, since $p(f(v))$ is unique for $f(v)$, implies that $f(v)$ is the only child of $f(u)$ (which is a reticulation node since $p(f(u))$ is not irreducible).\\
Suppose now that $u$ is a tree node, and let $v_1, v_2$ be its two children. Then, we know that $p(v_x) = p(f(v_x))$ for $x\in\{1,2\}$, and that  $p(f(u)) = y + p(f(v_1))p(f(v_2))$. Since each node is uniquely characterized by its polynomial, it means that both $f(v_1)$ and $f(v_2)$ are strong descendants of $f(u)$. By an argument analogous to that in the proof of Lemma \ref{lem:diff-equiv}, we can deduce that $f(v_1)$ and $f(v_2)$ are actually the children of $f(u)$.\\
Now, we prove that $f$ preserves the labels on the leaves and on the reticulations. If $u\in L(N_1)$, then $f(u)\in L(N_2)$. Since $u \in L(N_1)$, by definition,  $p(u)=\varphi_1(u)$. Moreover, $p(u)=p(f(u))$ because leaves are tree nodes. Since $f(u) \in L(N_2)$,  $p(f(u))=\varphi_2(f(u))$. Then, $\varphi_1(u)= \varphi_2(f(u))$. Now, let $u \in R(N_1)$ (a reticulation on $N_1$). By definition, $p(u)=\ell_1(u) p(v)$, where $v$ is the single child of $u$. We have seen above that $p(f(u))=\ell_1(u)p(f(v))$; but, since $f(u)$ is a reticulation in $N_2$ and $f(v)$ is its single child, by definition, $p(f(u))=\ell_2(f(u))p(f(v))$. Then, $\ell_1(u)=\ell_2(f(u))$. 
\end{pf}

\begin{thm}\label{thm:sep-iso}
Let $N_1, N_2$ be two internally labelled separable phylogenetic networks. If they are equivalent \emph{modulo} strong paths, then they are isomorphic.
\end{thm}
\begin{pf}
By Lemma \ref{lem:sep-diff}, if $N_1$ and $N_2$ are separable, then $p(u_1) = p(u_2)$ implies $u_1=u_2$ for any internal node in either $N_1$ or $N_2$. Then, if we are able to find a bijection $f$ between the sets of nodes satisfying the premises of Lemma \ref{lem:iso}, we will be able to apply it and show the result.\\
Now, $N_1$ and $N_2$ are equivalent \textit{modulo} strong paths, and that means that there exists a bijection $f$ between the sets of tree nodes such that, for a fixed permutation $\sigma$ between the sets of labels with $\sigma(X) = X$, $p(u) =\ ^\sigma p(f(u))$ for any tree node $u$,  and if $u, v$ are tree nodes and $v$ is a strong descendant of $u$, then $f(v)$ is a strong descendant of $f(u)$. We shall show that this $f$ induces our bijection if we generalize it to any internal node (i.e., if we define it correctly over the reticulation nodes in $N_1$). In order to ease the notation, and without loss of generality, let $\sigma$ be the identity. \\
Let $v$ be a reticulation node in $N_1$, and $u$ a tree node that is a strong ancestor of it. Let $v^{(1)}, v^{(2)}$ be the children of $u$, and suppose that $v$ strongly descends from $v^{(1)}$. Let $w_1, w_2$ be the two (possibly equal) tree nodes that strongly descend from $u$.\\
Since $N_1$ is separable, in particular $u$ is separable, and we know that we can write $p(v^{(1)}) = \mu_1^{(1)} \ldots \mu_{r_1}^{(1)} p(w_1)$ and $p(v^{(2)}) = \mu_1^{(2)}\ldots\mu_{r_2}^{(2)} p(w_2)$. Now, by Lemma \ref{lem:entry}, either (1) there exists a tree node $u'$ that enters the neighbourhood of $u$ at $v$, or (2) it does not and both parents of $v$ are strong descendants of $u$.

Thus, we distinguish the following cases:
\begin{itemize}
    \item[(1)] There exists a tree node $u'$ that enters the neighbourhood of $u$ at $v$, and
    \begin{itemize}
        \item if $v$ is the only reticulation node at which $u'$ enters the neighbourhood of $u$ (that is $u'\in U_1^{(1)}(u)$), then $p(v) = \mu_{i_1}^{(1)}\ldots \mu_{r_1}^{(1)}p(w_1)$, where $\mu_{i_1}^{(1)},\ldots,\mu_{r_1}^{(1)}$ are the only polynomials in $\lambda_1,\ldots,\lambda_r$ that divide both $p(u)-y$ and $p(u')-y$.
        \item if $u'$ also enters the neighbourhood of $u$ at some $v'$ and there is no strong path between $v$ and $v'$ (that is $u'\in U_3(u)$), then $p(v) = \mu_{i_1}^{(1)}\ldots \mu_{r_1}^{(1)}p(w_1)$, where $\mu_{i_1}^{(1)},\ldots,\mu_{r_1}^{(1)}$ are the only polynomials in $\lambda_1,\ldots,\lambda_r$ that divide both $p(u)-y$ and $p(v^{(1)})$.
        \item if $u'$ also enters the neighbourhood of $u$ at some $v'$ that is a strong ancestor of $v$ (that is a case where $u'\in U_2^{(1)}(u)$), then $p(v) = \mu_{i_1}^{(1)}\ldots \mu_{r_1}^{(1)}p(w_1)$, where $\mu_{i_1}^{(1)},\ldots,\mu_{r_1}^{(1)}$ are the only polynomials in $\lambda_1,\ldots,\lambda_r$ such that they divide $p(u)-y$ and, for every $j\in \{i_1, \ldots, r_1\}$, $(\mu_{j}^{(1)})^2\mid p(u')-y$.
        \item if $u'$ also enters the neighbourhood of $u$ at some $v'$ that is a strong descendant of $v$ (that is a case where $u'\in U_2^{(1)}(u)$), then $p(v) = \mu_{i_1}^{(1)}\ldots \mu_{r_1}^{(1)}p(w_1)$, where $\mu_{i_1}^{(1)},\ldots,\mu_{r_1}^{(1)}$ are the only polynomials in $\lambda_1,\ldots,\lambda_r$ that divide both $p(u)-y$ and $p(u')-y$.
    \end{itemize}
    Notice that the above arguments are independent of whether $w_1 = w_2$ or not.
    \item[(2)] Both parents of $v$ are strong descendants of $u$ (and so $w_1 = w_2$). Let $\mu_{i_1}$ the label of the reticulation $v$ and let $\mu_{i_1},\ldots, \mu_{r_3}$ the labels of reticulations in the strong path from $v$ to $w_1$. Then $p(v) = \mu_{i_1} \ldots \mu_{r_3} p(w_1)$, where $\mu_{j}$ for $j\in\{i_1,\ldots, r_3\}$ are the only polynomials in $\lambda_1, \ldots, \lambda_r$ such that $(\mu_{j})^2\mid p(u)-y$.
\end{itemize}

Since $N_2$ is also separable, in particular $f(u)$ is separable, and since $p(f(u)) = p(u)$ (because $N_1$ and $N_2$ are equivalent \textit{modulo} strong paths), some of its children cannot be a tree node. Therefore, if $v_*^{(1)}, v_*^{(2)}$ are its children, there must exist a tree node $u_1$ that is either a strong ancestor of $v_*^{(1)}$ but not of any other strong descendant of $f(u)$ or a strong ancestor of, say, $v_*^{(1)}$ and of one of its strong descendants. This node will allow us to characterize $p(v_*^{(1)})$. But since $N_1$ and $N_2$ are equivalent \textit{modulo} strong paths, there exists $f^{-1}(u_1)$ in $N_1$ that satisfies the same condition with regard to the pair $u, v^{(1)}$ in $N_1$, and so $p(v^{(1)})=p(v_*^{(1)})$ and $p(v^{(2)}) = p(v_*^{(2)})$. Thus, we set $f(v^{(1)}) = v_*^{(1)}$ and $f(v^{(2)}) = v_*^{(2)}$.\\
Now, for any $v_*$ reticulation node strongly descending from either $v_*^{(1)}$ or $v_*^{(2)}$, any of its strong ancestors that are tree nodes are such that there exists a tree node in $N_1$ with its same polynomial (and thus, is a strong ancestor of some $v$ strongly descending from $u$). Therefore, we will have that $p(v) = p(v_*)$, and we can then set $f(v) = v_*$.
\end{pf}

Theorem \ref{thm:imlt-inj} and Theorem \ref{thm:sep-iso} together imply the following  main result.

\begin{thm}\label{teo:pol-iso}
Let $N_1, N_2$ be two internally labelled separable phylogenetic networks, and $\sigma$ a permutation of their labels such that $\sigma(X) = X$. If $p(N_1) =\ ^\sigma p(N_2)$, then $N_1$ and $N_2$ are isomorphic.
\end{thm}

\subsection{Orchard networks}

In this subsection we prove that the phylogenetic networks in the class of orchard networks \cite{erdos2019class} are separable. These (strictly) include tree-child networks.

Before we recall the definition of orchard networks, we need to  introduce some definitions.
Let $N$ be a phylogenetic network on $X$. Let $\{a, b\} \subseteq X$. The set $\{a, b\}$ is a \emph{cherry} of $N$ if $a$ and $b$ share a parent.  Let $p_a$ and $p_b$ the parents of $a$ and $b$, respectively. If $p_b$ is a reticulation and $(p_a, p_b)$ is
an arc in $N$, then $\{a, b\}$ is a \emph{reticulated cherry} of $N$. \\
Let $N$ be a phylogenetic network and let $\{a, b\}$ be a cherry of $N$. Then ``\emph{reduce} $b$'' is the operation of deleting $b$ and suppressing the resulting elementary node. If $p_a=p_b$ is the root of $N$, then delete $b$ and the root. If $\{a, b\}$ is a reticulated cherry of $N$ in which $p_b$ is the reticulation, ``\emph{cut} $\{a, b\}$'' is the operation of deleting $(p_a, p_b)$, and suppressing the two resulting elementary nodes. For both operations, we say that a \emph{cherry-reduction} is performed on $N$. 

Let $N$ be a phylogenetic network. The sequence $N=N_0, N_1, \ldots, N_k$  of phylogenetic networks is a \emph{cherry-reduction sequence} of $N$ if, for all $i \in \{1,\ldots,k\}$, the phylogenetic network $N_i$ is obtained from $N_{i-1}$ by a (single) cherry-reduction. 
Then, a phylogenetic network $N$ 
is \emph{orchard} if there exists a cherry-reduction sequence $N=N_0, N_1, \ldots, N_k$ of $N$ such that $N_k$ consists of a single vertex.

\begin{thm}\label{thm:sep-orch}
Orchard networks are separable.
\end{thm}
\begin{pf}
Let $N$ be an orchard network and let $N=N_1,\ldots,N_k$ be a sequence of cherry-reductions of $N$. We prove that, for any $i\in\{1,\ldots,k-1\}$, if $N_i$ is not separable, then $N_{i+1}$ is not either. This means that if $N$ is not separable, the last network in every  cherry-reduction sequence cannot be a single vertex, reaching a contradiction due to $N$ being orchard. 
\\
If a reduction of a leaf in a cherry is produced there is nothing to prove because it does not involve reticulation nodes. Then suppose that a cut of a reticulated cherry $\{a, b\}$ is produced in $N_i$. Let $p_a$ and $p_b$ the parents of $a$ and $b$, respectively, and let $p_b$ the reticulation node. Then $p_a$ is a tree node. Moreover $p_a$ is a separable node in $N_i$ because the single strong descendant that is a reticulation node of $p_a$ is $p_b$. Then, $N_i$ is not separable due to some other tree node. 
\\
Notice that the cut of the reticulated cherry $\{a,b\}$ does not change the relation of strong descendance in the remaining nodes; i.e., $u, v$ were such that $v$ strongly descended from $u$ in $N_i$ if, and only if, the correspondent nodes in $N_{i+1}$ satisfy this condition too. More precisely, let $u$ be a non separable tree node, $v^{(1)}, v^{(2)}$ its children and $w_1, w_2$ the tree nodes that strongly descend from it. By Remark \ref{rem:not-sep} this means that, to begin with, neither $v^{(1)}$ nor $v^{(2)}$ are tree nodes and, if $w_1 \neq w_2$, all the strong ancestors of $v^{(1)}, v^{(2)}$ that are not $u$ are in $U_3(u)$. Now, $p_a$ can never be in $U_3(u)$ because one of its children is a leaf, $a$. Therefore, the cut of the reticulated cherry $\{a, b\}$ would not affect the non separability of $u$. Suppose now that $w_1 = w_2$. By Remark \ref{rem:not-sep}, if $v$ is the first reticulation node that is strong descendant of both $v^{(1)}, v^{(2)}$, the reticulation node $p_b$ cannot be in the strong paths from $v^{(1)}$ to $v$ and from $v^{(2)}$ to $v$ (note also that must be $p_b \neq v$). Then, both strong paths remain untouched to the cut of the reticulated cherry and also the set of strong ancestors of $v^{(1)}$ and $v^{(2)}$ that cause the non separability of $u$. Therefore, any non separable tree node in $N_i$ continues to be so in $N_{i+1}$.\end{pf}

\subsection{Unlabelled version}

Throughout this paper we have not made any use of the different labels of the leaves of an IMLN, and so the arguments could be translated, \textit{mutatis mutandis}, to IMLN's whose leaves are not labelled (although internal labels would still be necessary), modelled by labelling all leaves using a single variable $x$, to give a polynomial in $\ZZ[x,\lambda_1,\dots,\lambda_r,y]$. Again, for the case of phylogenetic networks, this would require that given two unlabelled phylogenetic networks we consider internally labelled phylogenetic networks with the same topology. This leads to the following proposition:
\begin{prop}\label{prop:unlabelled}
Let $N_1, N_2$ be two internally labelled separable phylogenetic networks whose leaves are all labelled by $x$. Then, $p(N_1) = p(N_2)$ implies that $N_1$ and $N_2$ are isomorphic.
\end{prop}


\section{Conclusion}\label{s:conc}
In this paper a new complete polynomial invariant for a class of (binary) phylogenetic networks, that of separable networks, is introduced. It generalizes results in both \cite{liu2020polynomial} for phylogenetic trees and in \cite{janssen2021comparing} for phylogenetic networks where their set of embedded spanning trees (like tree-child) characterizes it. The introduced polynomial $p$ is a generalization of the Liu polynomial and it is defined in a more generic structure of networks, called IMLN's, where the reticulations are also labelled with labels other than those on the leaves. We prove that  for the case of separable phylogenetic networks, the internally labelled structure derived from those is completely characterized by the polynomial. This induces a complete polynomial invariant for separable phylogenetic networks. That is, given two separable phylogenetic networks $N_1$ and $N_2$ on $X$, we could fix an internally labelled phylogenetic network from it, say $N_1^*$,  by bijectively labelling the reticulations. Then, if we consider all possible internally labelled phylogenetic networks obtained from $N_2$ by the permutation of all its variables,  $X$ and the reticulations, we can compare $p(N_1^*)$ with the polynomial of all the networks obtained from $N_2$.  Note that, due to Proposition \ref{prop:unlabelled}, we could avoid the permutation of the labels on $X$, reducing the cost of this computation.

Establishing a complete polynomial  invariant for phylogenetic networks opens the door to several interesting opportunities for exploration, such as new ways to define metrics on networks, 
fast methods to distinguish networks, and possibly ways to extract important features of a network by examining this polynomial.  To this end, it may be helpful to understand whether a particular polynomial is derived from a network or not (for clearly not all irreducible polynomials give networks).

Furthermore, the computation of $p(N)$ here may be performed reticulation-by-reticulation for some network classes, eg orchard networks~\cite{erdos2019class}. That is, suppose that $N$ is an internally labelled phylogenetic network derived from an orchard network and $N=N_0, N_1, \ldots, N_k$ is a complete cherry reduction sequence of $N$ (that is $N_k$ is a single node). We can perform an assignment of polynomials to all leaves in every intermediate IMLN $N_j$. Finally, $p(N)$ is the polynomial assigned to the single node in $N_k$. Start by assigning $p(u)=\varphi(u)$, for every leaf $u$ in $N_0$. Then, let $\{v_1,v_2\}$ be the two leaves involved in the cherry-reduction to move from $N_j$ to $N_{j+1}$ and let $p(v_i)$ be the polynomial assigned to $v_i$ in $N_j$ for $i\in \{1,2\}$.  Then, 

\begin{itemize}
    \item if $\{v_1,v_2\}$ is a cherry, assign to the resulting leaf in $N_{j+1}$ the polynomial  $y+p(v_1)p(v_2)$.
    \item if $\{v_1,v_2\}$ is a reticulated cherry (being $v_2$ the child of the reticulation labelled by $\lambda_i$), assign to the resulting leaf in $N_{j+1}$ coming from the parent of $v_1$ the polynomial $y+\lambda_i p(v_1)p(v_2)$, and to the resulting leaf in $N_{j+1}$ coming from the parent of $v_2$, the polynomial $\lambda_ip(v_2)$.
\end{itemize}

It would be interesting to investigate more  optimisations for general or for specific subclasses of phylogenetic networks.

\begin{figure}[!ht]
    \centering
 \includegraphics[scale=0.5]{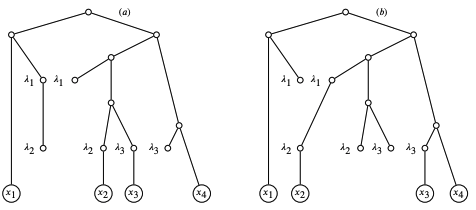}  
     \caption{{\bf Subtrees of $U(N)$}. Let $N$ be the internally labelled phylogenetic network depicted in Fig \ref{fig:unf}. The figure depicts two (IMLT) subtrees of $U(N)$.}
    \label{fig:newick}
\end{figure}

It would also be interesting to think about ways to reduce the complexity of the polynomial assigned to a network; even at the expense of a loss of the uniqueness of this assignment. One possibility would be, for instance, to define a polynomial for a phylogenetic network over the IMLT into which is transformed the network following a similar approach that allow the computation of its extended Newick format \cite{cardona2008extended}. Consider, for example, this: for every reticulation, split it (also copying its label) in two copies, the first such copy with one of its parent and its child, and the other copy with the other parent and no children. See two examples of this decomposition in Fig~\ref{fig:newick} from the internally labelled phylogenetic network $N$ depicted in Fig~\ref{fig:unf}. Clearly, this transformation process is not unique, and different IMLT's can be obtained from the same network; but different networks result in disjoint sets of IMLT's. Notice that this process can be understood as a way to prune irrelevant subtrees of the IMLT $U(N)$, with the goal to keep enough information to code the network. Roughly speaking, to recover the network from these IMLT's one should only merge every pair of nodes labelled by the same $\lambda_i$. 
Applying the definition of the polynomial $p$ to these IMLT's, we obtain, for the example depicted in Fig~\ref{fig:newick} (a), the polynomial 
\begin{equation*}
\begin{split}
p(N) = & y+ y^2 + y^3 +  \lambda_1y^3 + \lambda_1\lambda_3x_4y^2  +
 \lambda_1\lambda_2\lambda_3x_2x_3y^2 + \lambda_1\lambda_2\lambda_3^2x_2x_3x_4y + \\ & + \lambda_1\lambda_2x_1y + \lambda_1\lambda_2x_1y^2 + \lambda_1\lambda_2\lambda_3x_1x_4y  + \lambda_1^2\lambda_2x_1y^2 + \lambda_1^2\lambda_2\lambda_3x_1x_4y + \\ & +
 \lambda_1^2\lambda_2^2\lambda_3x_1x_2x_3y + 
 \lambda_1^2\lambda_2^2\lambda_3^2x_1x_2x_3x_4,
\end{split}
\end{equation*}
where (some of) the terms are notably simpler than in the original.

There are potentially many further questions arising that relate to phylogenetic networks more broadly.  For instance, do embedded spanning trees characterize general internally labelled phylogenetic networks? That is, if we keep the labels on elementary nodes (which come from reticulation nodes) of the embedded spanning trees, can we extend the results in~\cite{francis2018identifiability} from tree-child networks to more general networks? Which classes of phylogenetic networks are separable? Do \textit{FU}-stable networks require all the labels of the polynomials $\lambda_1,\ldots, \lambda_r$ or can these be replaced by a single variable $\lambda$? And, over all, is there a complete characterization in topological terms of the phylogenetic networks that are characterized by the polynomial introduced in this article?

With all this, we hope that the results here will stimulate these and many other investigations.

\section*{Acknowledgment} 
JCP and TMC were supported by the Ministerio de Ciencia e Innovación (MCI), the Agencia Estatal de Investigación (AEI) and the European Regional Development Funds (ERDF); through project PGC2018-096956-B-C43 (FEDER/MICINN/AEI).

The authors thank Francesc Rosselló for helpful comments and suggestions.

\bibliographystyle{alpha}
\bibliography{mybibfile}

\end{document}